\documentclass[onecolumn,amsmath,amssymb,aps,nopacs,superscriptaddress]{revtex4}

\usepackage[dvips]{graphicx}
\usepackage[dvips]{color}
\usepackage{hyperref,breakurl,amsmath,amssymb,pst-node,eepic}
\usepackage{graphicx}% Include figure files
\usepackage{dcolumn}% Align table columns on decimal point
\usepackage{bm}% bold math
\usepackage{hyperref}

\begin{document}

\title{Offline Biases in Online Platforms: a Study of Diversity and Homophily in Airbnb}

\author{Victoria Koh}
\affiliation{Department of Computer Science, University College London, 66-72 Gower Street, WC1E 6EA London, UK}
\author{Weihua Li}
\affiliation{Department of Computer Science, University College London, 66-72 Gower Street, WC1E 6EA London, UK}
\affiliation{Systemic Risk Centre, London School of Economics and Political Science, Houghton Street, WC2A 2AE London, UK}
\author{Giacomo Livan}
\affiliation{Department of Computer Science, University College London, 66-72 Gower Street, WC1E 6EA London, UK}
\affiliation{Systemic Risk Centre, London School of Economics and Political Science, Houghton Street, WC2A 2AE London, UK}
\author{Licia Capra}
\affiliation{Department of Computer Science, University College London, 66-72 Gower Street, WC1E 6EA London, UK}

\date{\today}

\begin{abstract}
How diverse are sharing economy platforms? Are they fair marketplaces, where all participants operate on a level playing field, or are they large-scale online aggregators of offline human biases? Often portrayed as easy-to-access digital spaces whose participants receive equal opportunities, such platforms have recently come under fire due to reports of discriminatory behaviours among their users, and have been associated with gentrification phenomena that exacerbate preexisting inequalities along racial lines. In this paper, we focus on the Airbnb sharing economy platform, and analyse the diversity of its user base across five large cities. We find it to be predominantly young, female, and white. Notably, we find this to be true even in cities with a diverse racial composition. We then introduce a method based on the statistical analysis of networks to quantify behaviours of homophily, heterophily and avoidance between Airbnb hosts and guests. Depending on cities and property types, we do find signals of such behaviours relating both to race and gender. We use these findings to provide platform design recommendations, aimed at exposing and possibly reducing the biases we detect, in support of a more inclusive growth of sharing economy platforms.
\end{abstract}

\maketitle

\section{Introduction}

Sharing economy platforms are new manifestations of century old phenomena. Resource circulation systems that facilitate the exchange of underutilized goods or services between consumers have long existed in the form of flea markets, garage sales, second-hand shops, just to name a few. However, what used to be small scale and local instances of collaborative consumption, have now become massive online marketplaces, where face-to-face interactions have been replaced by technology-mediated ones \cite{zloteanu2018plos}. 

A fundamental question arises about the role that such decentralized, largely unregulated, online platforms play in our societies. Often thought of as level-playing fields, where all participants receive the same opportunities, sharing economy platforms might instead end up acting as {\em online} aggregators of well-known {\em offline} human dynamics and biases. Indeed, a number of studies have suggested that some of the big sharing economy players are acting as accelerators of gentrification pheonomena that are already underway in large cities. For example, Airbnb has led to the emergence of short-term rent gaps between different areas of New York City \cite{wachsmuth2018airbnb} and has contributed to exacerbating the affordable housing crisis in Los Angeles \cite{lee2016airbnb}. These phenomena, in turn, typically accelerate preexisting divides along racial lines, fostering inequalities between the Airbnb community and the communities living in the neighbourhoods where Airbnb has a significant presence (see, e.g., \cite{gentri}).

Moreover, sharing economy platforms have come under multiple allegations over discrimination episodes taking place within the platforms themselves. For example, Uber drivers were found to be twice more likely to cancel trips requested by passengers with African-American sounding names compared to White-sounding names (even though Uber penalizes drivers for cancellations) \cite{ge2016racial}. Similarly, Airbnb's hosts were found to be turning down potential guests based upon their racial background \cite{edelman2017racial}. 

While headway has been made to tackle the unintended consequences brought about by sharing economy platforms, the debate on their socio-economic impact is still in its infancy, and only relies on a handful of studies or on anecdotal evidence. This, in turn, delays the execution of targeted interventions to expose, and possibly reduce, such consequences. The goal of this paper is to contribute to inform such a debate by performing a large scale empirical analysis aimed at detecting systematic statistical evidence of `offline' biases taking place in online sharing economy platforms.

We study the Airbnb hospitality service, and focus first on the composition of its user base, with the aim of assessing its diversity both in general terms and then contextually with respect to the city hosting it. Second, we employ a network methodology to assess the statistical significance of host-guest interactions in Airbnb. In particular, 
we focus on {\em homophily}, i.e. the social phenomenon where people gravitate towards those like themselves \cite{mcpherson2001birds}, and on its opposite, heterophily. We also study the tendency to avoid members of a social group with different social traits, which we refer to as {\em avoidance}. While avoidance is universally deemed as unacceptable, homophily has sometimes been perceived as `natural', and thus judged in a more accepting way. However, several studies have shown that the aggregation of slightly biased individual preferences can lead to unintended and collectively undesirable consequences, as evidenced by Schelling's work on urban racial segregation \cite{schelling1969models,schelling1971dynamic}, and Neal's work on school children's development \cite{neal2010}.

In performing this study, we make three contributions:

\begin{itemize}
\item We gather data about Airbnb hosts, guests, and their interactions for five cities, spanning three different continents (Airbnb Data section). These are Amsterdam (The Netherlands), Dublin (Ireland), Hong Kong (China), Chicago and Nashville (U.S.). We have chosen them so to cover geographically (and culturally) different cities, as well as to cover variances in size, population composition, and cost of living.

\item We study the diversity of the Airbnb user base in the above five cities along the dimensions of gender, age, and race. We find the Airbnb community to be predominantly female, and overwhelmingly young and White. In line with the aforementioned literature, we find the majority of hosts to be White even in cities whose racial composition is significantly more diverse (Results section).

\item We model Airbnb's peers and interactions as nodes and edges in a bi-partite graph, and use a statistical method based on network rewiring to systematically identify edges (i.e., guest-host pairings) that cannot be attributed to chance (Method section). We apply such a method to the five cities under study, and, depending on the specific city and property type, find signals of homophily, heterophily and avoidance. We find such signals to be rather strong in the case of gender, rather weak (although still statistically significant) in the case of race, and mostly absent in the case of age (Results section).

\end{itemize}

These results echo other findings in the literature (see next section), and provide concrete evidence about how sharing economy platforms are being appropriated in different city contexts, possibly resulting in large divides between the online communities who can enjoy the benefits of the sharing economy and the `offline' urban communities who are most exposed to its expansion. They also offer an opportunity to inform the design of tailored technology interventions aimed at exposing, and possibly reducing, certain behaviours, while also providing the means to monitor their effects (Discussion section).

\section{Related Work}

Upon its inception, the Internet was expected to create a global level playing field, where the inequalities of the `offline' world would be overcome thanks to easy access to digital opportunities. Yet, reality has been very different. As it is invariably the case, different social groups are not equally equipped to face technological innovation in its early stages, which typically exacerbates preexisting inequalities \cite{aghion2015innovation}.

The sharing economy, as a whole, has been no exception. Indeed, a handful of studies have shown that the ability to seize the sharing economy's opportunities is often severely limited by geographical and socio-economic constraints. For example, Airbnb listings are usually more concentrated in wealthier, more attractive areas populated by young and tech-savvy residents \cite{quattrone2016benefits}. Similarly, TaskRabbit users from areas with low socioeconomic status and/or low population density were found to have a harder time both when selling their services and when seeking to outsource work to potential taskers \cite{thebault2015avoiding}, while individuals living in deprived Chicago suburbs have been found to have a harder time to get an Uber ride \cite{thebault2017towards}.

Similarly, the Internet's promise to circumvent physical barriers and improve communication between social groups has not always been upheld. For example, episodes of racial discrimination in online social networks have been extensively documented \cite{tynes2011racial,daniels2009cyber}. Also, a vast amount of scholarly work has been devoted to understanding the formation of online preferential relationships between individuals. This has often been explained either in terms of interest-based homophily, e.g., showing the impact of ideological homophily in determining the opinions and content individuals are exposed to on social media \cite{bakshy2015exposure}, or in terms of homophily driven by {\em demographics}.

Studies of early social networks, e.g., MySpace \cite{thelwall2009homophily}, have identified race, gender, and age as the main demographic features driving online homophily, and such elements kept recurring in more recent studies. Indeed, evidence of racial \cite{wimmer2010beyond} and gender \cite{bamman2014gender} homophily has been reported in Facebook and Twitter, respectively, and evidence of both has been documented in the social networks underpinning location sharing applications \cite{guha2015birds}. Age homophily is somewhat less studied, but still documented in a study of the Facebook social graph \cite{ugander2011anatomy} and in niche environments such as virtual worlds \cite{huang2009virtually,foucault2010friend}.

Our work follows this stream of literature and investigates  whether well known `offline' biases also take place in sharing economy platforms. A handful of recent studies have started to look at such platforms from this perspective. Indeed, recently published work \cite{ge2016racial} found evidence of both gender and racial discrimination in Uber and Lyft, as female passengers were disproportionally taken on longer and more expensive routes, while passengers with African American-sounding names were twice as likely to receive trip cancellations from Uber drivers compared to passengers with White-sounding ones (even though Uber penalizes drivers for cancellations). Similarly, another study \cite{hannak2017bias} found gender and race to have an impact on worker evaluations in online freelance marketplaces.

Evidence of biased behaviour was also found in Airbnb by means of a field experiment \cite{edelman2017racial}. In particular, guests were found to be 16\% less likely to have their booking accepted if they had a distinctly African American-sounding name when compared to identical guests with White-sounding names instead. Similarly, in \cite{edelman2014digital} it was found that non-Black hosts charged on average 12\% more for an equivalent rental compared to Black hosts, and similar results were replicated in a subsequent study on Airbnb \cite{kakar2016effects}, where Asian and Hispanic hosts were found to rent at prices 9.3\% and 9.6\% lower, respectively, than their White counterparts.

While the above works investigate some specificities of user demographics and interactions in sharing economy platforms, a systematic analysis of these dimensions across the fundamental features of gender, age, and race is still lacking. This work aims at filling this void, by providing ($i$) an overview of the composition and diversity of Airbnb's community, and ($ii$) a quantitative method to dissect the anatomy of user-user interactions in sharing economy platforms (and Airbnb in particular), providing statistical evidence of homophily and avoidance between certain user groups.

\section{Airbnb Data}
\label{sec:airbnb_data}

In order to perform this study, we needed two types of data: demographic characteristics of hosts and guests (i.e., gender, age, race); and their pairing dynamics (i.e., who stayed with whom). Since we hypothesise that peers' behaviours might vary in different geographic (and cultural) contexts, we chose to perform this study on a per city level, rather than treating the whole of Airbnb as a single analytical context.

To begin with, we accessed city snapshots that the website InsideAirbnb\footnote{{http://insideairbnb.com/get-the-data.html}} already makes available. We chose five cities (Amsterdam, Chicago, Dublin, Hong Kong, Nashville) so to have high geographic diversity (these cities span three different continents), as well as high diversity in terms of population composition and cost of living. Records of Airbnb hosts, guests and stays go from 2008 to 2016 for all cities except Nashville, whose Airbnb records start in 2009. For each city, InsideAirbnb makes available a full list of host IDs (from their `listings' file). We used these IDs to query the Airbnb website and further acquire a host profile picture, the type(s) of property they were renting out (i.e., full property, private room in a shared property, or shared room), and the full list of IDs of all guests that ever left a review to such host (and for what property). We then further queried the Airbnb website with the guest IDs to acquire their profile pictures. Since Airbnb does not explicitly make available a peer's gender, age and race as attribute-value pairs in the peers' profile, we used image processing software on the collected profile pictures to automatically extract this information. In particular, we first used face localisation software to detect whether the profile picture contained a human face, and if so, to identify the portion in the picture containing it. We tested both FaceReact\footnote{{https://github.com/cooyzee/face-react}} and Indico\footnote{{https://indico.io/}} on a manually curated sample of 50 Airbnb images, so to contain a mix of pictures with and without human faces, and with and without background clutter. We found Indico to be significantly more accurate, especially for human images taken at an angle rather than straight-facing the camera. We thus continued only with the latter. Having extracted the bounding box containing a human face, we then used face recognition software to extract attributes. We tested Betaface,\footnote{{https://www.betaface.com/wpa/}} Sightcorp F.A.C.E,\footnote{{https://face-api.sightcorp.com/}} and Face++\footnote{{https://www.faceplusplus.com/}} on a subset of 250 Airbnb images. We found all three to be equally accurate when detecting gender. Sightcorp was found to be significantly more reliable in recognising age groups, and Betaface in extracting race (note that our analyses will focus exclusively on race, not on ethnicity; in particular, we will focus on three main race categories, i.e., White, Black, and Asian). We thus worked with Sightcorp and Betaface in parallel. We manually verified their accuracy on all 250 test images, and found the confidence levels reported by both products to be $0.3 \in [0,1]$ or higher on images annotated correctly. Hence, we kept such value as a threshold for the ensuing automatic annotation; furthermore, we only retained pictures for which both face recognition software  products agreed on both gender and ethnicity. To understand how robust our results are when varying facial annotation accuracy, we repeated all our analyses after $(i)$ increasing the above threshold to 0.5, and $(ii)$ manipulating the data by changing the race annotation on a random sub-sample of the images. The results obtained from such analyses are reported in Appendix \ref{app_robust}.

In terms of pairing dynamics, Airbnb does not make visible who stays with whom, nor whether a stay request has been refused or cancelled. However, what it does make visible are reviews that hosts and guests leave to one another. We use these as proxies for the actual pairing dynamics. Studies show that over 65\% of stays result in a guest review and 72\% result in a host review \cite{fradkin2017determinants}, so most stays are indeed captured by reviews. At present, it is not known whether those who do {\em not} leave reviews in Airbnb belong to specific users' groups; a past survey study of Tripadvisor reviews \cite{Gretzel2007} did find that certain age and gender groups were more vocal than others, and this might also be the case in this context. Although the method we present next is still applicable, the validity of some of our findings might be impacted, and we will come back to this when we discuss limitations and future work (Conclusion section).

Summary statistics about the number of hosts, guests and pairings that we collected and annotated for each city under study are reported in Table \ref{tab:annotation}. 

\begin{table}[!htb]
  \caption{Number of hosts, guests, and host-guest pairs annotated for each city analysed}
  \label{tab:annotation}
  \begin{tabular}{lccc}
    \hline
    City & \# Hosts & \# Guests & \# Host-Guest Pairs \\
    \hline
    Amsterdam & 2,369  & 69,923  & 71,779  \\
    Chicago & 1,706  & 21,105  & 22,493  \\
    Dublin & 1,039  & 2,618  & 2,785  \\
    Hong Kong & 1,233  & 12,103  & 13,330  \\
    Nashville & 630  & 1,712  & 2,017  \\
    \hline
  \end{tabular}
\end{table}

\section{Method}
We model Airbnb hosts and guests as nodes in a bi-partite graph, with a directed $g \rightarrow h$ edge with weight $w_{gh}$ representing the number of times guest $g$ stayed at host $h$. Since we hypothesise that pairing dynamics may vary across cities, as well as across type of rented property (full property vs. shared -- the latter comprising both private and shared rooms), we create and analyse a total of $10$ ($5$-cities $\times \ 2$-property types) bipartite networks.

Each such network is  analysed using a statistical {\em rewiring} approach designed to assess the significance of pairing patterns in each of the cities studied. More precisely, the method starts from a null hypothesis that a given guest-host pairing occurred randomly. It then proceeds to verify whether this hypothesis holds by creating ensembles of null network models through the rewiring of the original networks' edges, and by comparing the properties of such null network model against those observed in the actual, empirical networks. Crucially, the procedure is designed to preserve the heterogeneity of the original networks, as it produces null network configurations where the number of stays that each guest and host have had are both kept intact, therefore preserving correlations between demographic features and activity on Airbnb. In the following, we describe the details of this methodology, and the rationale for adopting it.
            
\begin{figure}[h]
\includegraphics[scale=0.3]{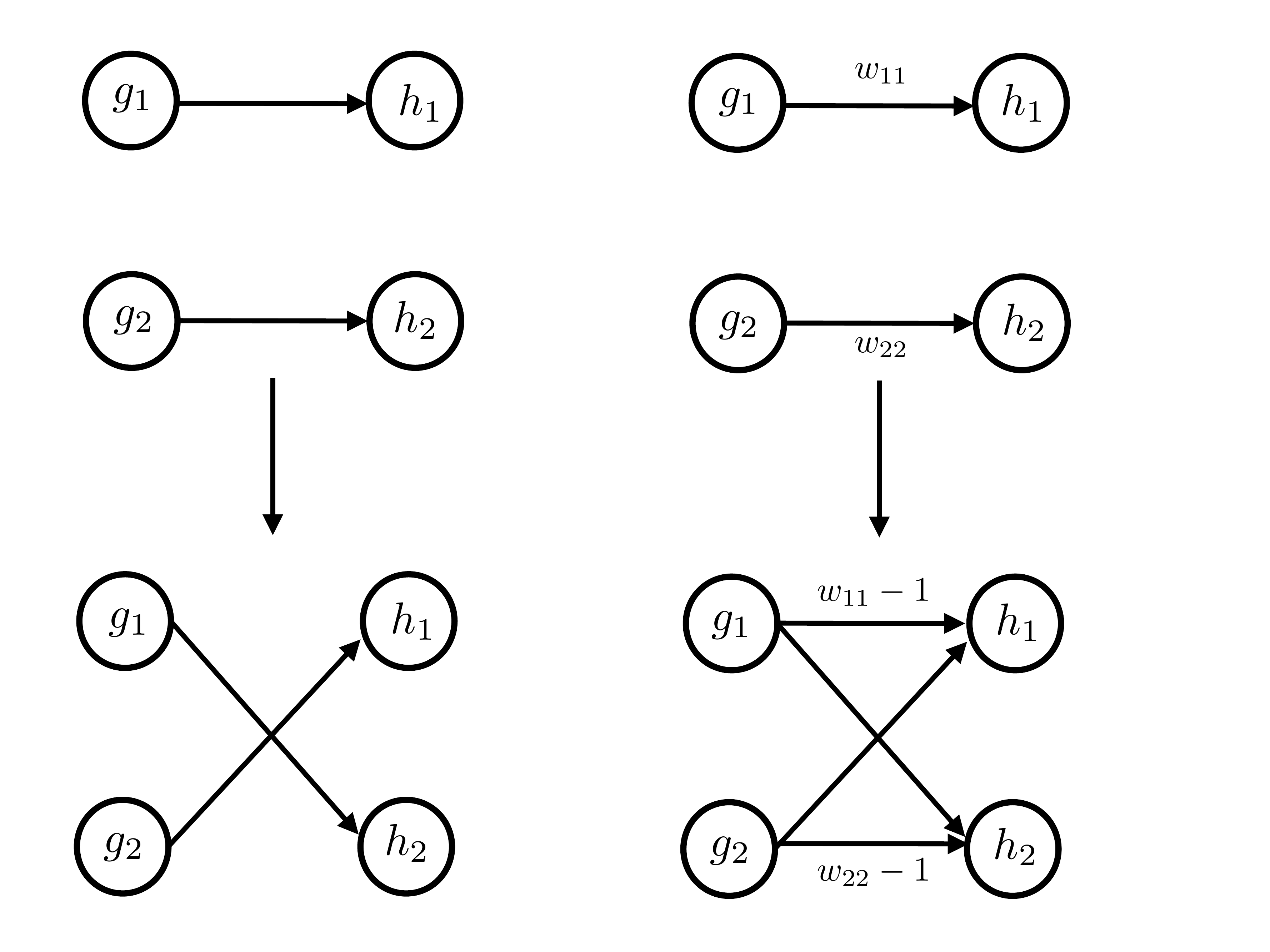}
\caption{xSwap rewiring moves. Left: swap of two links with unit weight. Right: swap of a unit weight subtracted from two links with weights larger than one.}
\label{fig:rewiring}
\end{figure}

Starting from an empirical bipartite network, we create a randomised version of it by iteratively performing xSwap operations  \cite{hanhijarvi2009randomization}. These amount to selecting two guest nodes ($g_1$ and $g_2$) and one of their corresponding host nodes ($h_1$ and $h_2$, respectively) at random, erasing the existing edges ($g_1 \rightarrow h_1$ and $g_2 \rightarrow h_2$) between both pairs, then reassigning them to each other ($g_1 \rightarrow h_2$ and $g_2 \rightarrow h_1$), as shown in Figure \ref{fig:rewiring} (left). Should either of the selected edges have a weight larger than one, the strength of the link is reduced by one and a unit weight is redistributed to the new host node (see Figure \ref{fig:rewiring} - right). These operations preserve both the outgoing weight of guest nodes and the incoming weight of host nodes. Therefore, repeated xSwap operations yield configurations which are exactly equivalent to the original networks in terms of their heterogeneity, but are instead fully randomized in terms of the relationships between their nodes \cite{livan2017excess}. In order to determine how many xSwap operations were needed before the rewired network configurations could be considered distinct enough (i.e., sufficiently uncorrelated) from their original counterparts, we computed the Kendall's Correlation Coefficient as suggested in \cite{bardoscia2013social}.

For each of the 10 empirical networks under consideration, we generated an ensemble of 1000 randomized network configurations with the above rewiring procedure, thus preserving the original networks' degree sequences. Since such a rewiring procedure generates null configurations that do not deviate substantially from the original networks themselves, we can resort to a fairly parsimonious numerical investigation based on such relatively small number of null configurations. On such configuration ensemble, we computed the frequency of interaction between different {\em groups} of hosts and guests based on gender, age, and race (e.g., female guests to male hosts, White guests to White hosts, etc.). We then took the 95\% confidence interval to define lower and upper bound probabilities for the occurrence of each pairing combination. This range represents the expected probabilities of pairing combinations taking place by chance. We then computed the pairing probabilities among the same groups in the original empirical network: should the actual values fall below or above the `by-chance' range, we take them to be statistically significant under/over-expressions of certain group interactions with respect to the null hypothesis. Note that the over-expression of interactions between two groups (or within a group) does not necessarily translate into the under-expression of interactions between other groups; we will use the former to detect a tendency of certain groups to {\em associate}, and the latter  to separately detect a tendency to {\em avoid} certain groups instead.

\section{Results}
In this section, we report the results obtained when applying the above method to analyse Airbnb users' pairing dynamics. Before doing so, we provide an overview of the demographic characteristics of Airbnb hosts and guests, whose profile pictures we scraped and annotated using the process described in the Airbnb Data section. We will discuss and elaborate on these results in the Discussion section.

\subsection{Demographics}
In Table \ref{tab:features_gender} we report summary statistics of Airbnb users in terms of {\em gender}, broken down by city and by property rental type (full vs shared). As shown, Airbnb has a predominantly female user base: in all cities under study, the net majority of both hosts and guests were found to be female, regardless of the property type, with such proportion getting close to or exceeding 60\% in a number of cases.

\begin{table}[!t]
  \caption{Airbnb host and guest population by gender (F=female)}
  \label{tab:features_gender}
%  \resizebox{\columnwidth}{!}{
  \begin{tabular}{lcc|cc}
    \hline
    & \multicolumn{2}{c|}{Full property rental} & \multicolumn{2}{c}{Shared property rental} \\
    \hline
    City & F Host & F Guest & F Host & F Guest \\
   \hline
    Amsterdam & 59\% & 55\% & 59\% & 58\%  \\
    Chicago   & 61\% & 58\% & 57\% & 56\%  \\
    Dublin    & 61\% & 59\% & 58\% & 54\%  \\
    Hong Kong & 56\% & 59\% & 60\% & 59\%  \\
    Nashville & 63\% & 66\% & 74\% & 59\%  \\
    \hline
  \end{tabular}
  %}
\end{table}         
             
\begin{table}[!t]
  \caption{Airbnb host population by race (W=White, A=Asian, B=Black)}
  \label{tab:features_ethnicity}
%  \resizebox{\columnwidth}{!}{
  \begin{tabular}{lccc|ccc}
    \hline
    & \multicolumn{3}{c|}{Full property rental} & \multicolumn{3}{c}{Shared property rental} \\
    \hline
    City & W & A & B & W & A & B \\
    \hline
    Amsterdam &  93\% & 5\%  & 2\% &  88\% &  10\% & 2\% \\
    Chicago   &  90\% & 8\%  & 2\% &  87\% & 10\% & 3\% \\
    Dublin    &  96\% & 4\%  & 1\% &  92\% &  7\% & 1\% \\
    Hong Kong &  64\% & 35\% & 2\% &  58\% & 40\% & 2\% \\
    Nashville &  92\% & 3\%  & 5\% &  92\% &  5\% & 2\% \\
    \hline
  \end{tabular}
  %}
\end{table}

We next consider {\em race}, once again as it varies by city and by property rental type (see Table \ref{tab:features_ethnicity}). We focus our attention on hosts only in this case, so to compare our results with available census information on the demographics of the cities under study. In all cities, the majority of hosts were found to be racially White, even in those with a markedly diverse racial makeup (e.g., Hong Kong, Chicago, Nashville). We shall comment extensively on this finding in the next section.

\begin{table}[!tb]
  \caption{Quintiles of Airbnb's host age distributions for full property rentals. $Q_1$ denotes the bottom $20\%$ of the age distribution, $Q_2$ denotes users falling between the bottom $20\%$ and $40\%$ of the age distribution, and so on.}
  \label{tab:age_groups}
 % \resizebox{\columnwidth}{!}{
  \begin{tabular}{lccccc}
    \hline
    & \multicolumn{5}{c}{Hosts} \\
    \hline
    City &  $Q_1$ & $Q_2$ & $Q_3$ & $Q_4$ & $Q_5$ \\
    \hline
    Amsterdam &  $<$ 26 & [26, 31) & [31, 34) & [34, 39) & $>$ 39 \\
    Chicago &  $<$ 26 & [26, 31) & [31, 35) & [35, 39) & $>$ 39 \\
    Dublin & $<$ 24 & [24, 29) & [29, 33) & [33, 38) & $>$ 38 \\
    Hong Kong & $<$ 27 & [27, 31) & [31, 35) & [35, 39) & $>$ 39 \\
    Nashville & $<$ 27 & [27, 31) & [31, 34) & [34, 39) & $>$ 39 \\    
    \hline
  \end{tabular}
  %}
\end{table}

The last demographic feature we consider is {\em age}. Table \ref{tab:age_groups} reports the breakdown of each considered city's host population based on quintiles (we found the summary statistics to be very similar in the case of hosts, as well as when separating hosts based on full or shared property rentals). As shown in Table \ref{tab:age_groups}, Airbnb is a rather `young' community, with the vast majority of users found to be between their mid-twenties and late thirties.

\subsection{Rewiring Analysis}
We now present results of Airbnb host/guest pairing dynamics and compare them with the results of the rewiring analytical method illustrated above. We break down results by focusing on one demographic characteristic at a time for both hosts and guests. In this work, we did not investigate interactions between different features (e.g., between age of a host and gender of a guest); the same methodology could be used in the future to study these interactions.

{\em Gender-related pairings.} As shown in Table \ref{tab:gender}, we found very different, strongly city-dependent patterns. Indeed, we found same gender interactions to be prevalent in some cities (e.g., Amsterdam) while in others we found interactions between hosts and guests of different genders to be prevalent (e.g, Nashville). Yet, these results \emph{per se} are not necessarily informative, as they partially echo each city's Airbnb population composition, and become statistically relevant only when compared with the \emph{expected} rate of interaction measured under the null hypothesis of random interaction encoded in the above link rewiring procedure. As reported in Table \ref{tab:gender}, depending on the city and property type we detect very different over/under-expression patterns. Namely, when looking at full property rentals, we find same-gender interactions to be over-expressed (homophily), and interactions between different genders to be under-expressed (avoidance), in all cities. Things change considerably when looking at shared properties, where in the case of Dublin we find interactions between different genders to be over-expressed (heterophily) and, symmetrically, same gender interactions to be under-expressed, while in Nashville we find interactions to be compatible with the null hypothesis.

%\begin{landscape}
%\begin{small}
\begin{table}
  \caption{Pairings between guest-host genders (F=female; M=male) in Airbnb. Values in brackets represent 95\% confidence level intervals obtained from the rewiring analysis, while the values to their right denote the corresponding empirically observed frequencies. Upward green (downward red) arrows highlight over-expressed (under-expressed) values.}
  \label{tab:gender}
\begin{center}
   \hspace{-2cm} \begin{tabular}{lclclclcl}
    \hline
    & \multicolumn{8}{c}{Full property rental} \\
    \hline
    City & FF & FF & FM & FM & MF & MF & MM & MM  \\
    \hline
    AMS & [29.39; 29.44]\% &  30.23\% $\color{green} \uparrow$ & [23.98; 24.03]\%  & 23.19\% $\color{red} \downarrow$ & [25.61; 25.65]\% & 24.82\% $\color{red} \downarrow$ & [20.93; 20.97]\% & 21.76\% $\color{green} \uparrow$ \\
    CHI & [33.57; 33.66]\% & 34.81\% $\color{green} \uparrow$ & [25.04; 25.13]\% & 23.89\% $\color{red} \downarrow$ & [23.60; 23.69]\% & 22.45\% $\color{red} \downarrow$ & [17.61; 17.70]\% & 18.84\% $\color{green} \uparrow$ \\
    DUB & [35.17; 35.40]\% & 37.80\% $\color{green} \uparrow$ & [25.29; 25.52]\% & 22.88\% $\color{red} \downarrow$ & [22.85; 23.08]\% & 20.44\% $\color{red} \downarrow$ & [16.23; 16.46]\% & 18.87\% $\color{green} \uparrow$ \\
    HK & [33.24; 33.33]\% & 33.79\% $\color{green} \uparrow$ & [24.51; 24.61]\% & 24.06\% $\color{red} \downarrow$ & [24.16; 24.27]\% & 23.72\% $\color{red} \downarrow$ & [17.89; 17.99]\% & 18.44\% $\color{green} \uparrow$ \\
    NAS & [41.29; 41.49]\% & 41.78\% $\color{green} \uparrow$ & [22.60; 22.80]\% & 22.32\% $\color{red} \downarrow$ & [23.14; 23.34]\% & 22.86\% $\color{red} \downarrow$ & [12.56; 12.76]\% & 13.05\% $\color{green} \uparrow$ \\
    \hline
    & \multicolumn{8}{c}{Shared property rental} \\  
    \hline    
        AMS & [28.55; 28.62]\% & 29.27\% $\color{green} \uparrow$ & [20.65; 20.72]\%  & 20.00\% $\color{red} \downarrow$ & [29.38; 29.44]\% & 28.72\% $\color{red} \downarrow$ & [21.28; 21.34]\% & 22.01\% $\color{green} \uparrow$ \\
        CHI & [31.50; 31.58]\% & 32.91\% $\color{green} \uparrow$ & [24.64; 24.72]\% & 23.32\% $\color{red} \downarrow$ & [24.53; 24.61]\% & 23.21\% $\color{red} \downarrow$ & [19.16; 19.24]\% & 20.57\% $\color{green} \uparrow$ \\
    	DUB & [30.88; 31.47]\% & 29.59\% $\color{red} \downarrow$ & [26.59; 26.85]\% & 28.14\% $\color{green} \uparrow$ & [22.68; 22.94]\% & 24.24\% $\color{green} \uparrow$ & [19.32; 19.59]\% & 18.03\% $\color{red} \downarrow$ \\
    	HK & [33.25; 33.37]\% & 33.89\% $\color{green} \uparrow$ & [23.96; 24.07]\% & 23.44\% $\color{red} \downarrow$ & [24.79; 24.91]\% & 24.28\% $\color{red} \downarrow$ & [17.77; 17.88]\% & 18.40\% $\color{green} \uparrow$ \\
    NAS & [42.33; 42.60]\% & 42.55\%  & [30.91; 31.18]\% & 30.96\%  & [15.27; 15.54]\% & 15.32\%  & [10.95; 11.22]\% & 11.17\%  \\
    \hline 
  \end{tabular}
  \end{center}
\end{table}
%\end{small}
%\end{landscape}
             
{\em Race-related pairings.} Results capturing pairing dynamics in terms of race are shown in Tables \ref{tab:ethnicity_full} and \ref{tab:ethnicity_shared} for full and shared properties, respectively, once again broken down per city and per rental type. As a general observation, we can note that the signals we detect are not as strong as the ones detected for gender, with most over/under-expressions being of the order of a few fraction percentage points. Yet, strictly speaking, in most cases we detect an over-expression of White-to-White, Asian-to-Asian, and Black-to-Black guest-host pairings, regardless of whether the property was a shared rental or not (although with notable exceptions in Dublin and Nasvhille). Of note, the presence of racial homophily was found to be strongest in Hong Kong, both in terms of White-to-White and Asian-to-Asian guest-host pairings.

If we now change focus to pairing dynamics between different racial groups, we found results to be almost symmetric: that is, the  majority of pairings between hosts and guests from different racial groups were under-expressed, although with a few exceptions (see, e.g., Dublin's full property case), and these results appear to be largely independent from the property rental type. 

We verified the robustness of the above findings against possible confounding factors associated with the users' economic status. Namely, we checked whether the homophily and avoidance patterns shown in Tables \ref{tab:ethnicity_full} and \ref{tab:ethnicity_shared} might arise as a byproduct of correlations between racial background and wealth / income. We used the number of Airbnb properties owned and the price charged for a week-long stay as proxies for a host's income. We then performed two types of analysis: {\em (i)} a matched pair analysis \cite{ho2011matchit}, to measure the rate of interaction of White guests across groups of White and non-White hosts with similar levels of income; and {\em (ii)} a rewiring analysis on the sub-networks obtained after removing all hosts belonging to the top and bottom thirds of the income distribution. In both cases, by focusing our attention on hosts of similar economic status, we try to control for hosts' wealth. The results obtained from the matched pair analysis were statistically significant only in Hong Kong and Chicago (full properties); in these settings, they do confirm those reported in  Table \ref{tab:ethnicity_full}; results obtained from the rewiring analysis were all significant and fully in line with the ones reported in Tables \ref{tab:ethnicity_full} and \ref{tab:ethnicity_shared}. Full results are reported in Appendix \ref{app_econ}.
      
\begin{table}[!h]
 \small
  \caption{Pairings between racial backgrounds of Airbnb guests and hosts (W=White, A=Asian, B=Black) in full property rentals. Values in brackets represent 95\% confidence level intervals obtained from the rewiring analysis, while values below them denote the corresponding empirically observed frequencies. Upward green (downward red) arrows highlight over-expressed (under-expressed) values.}
  \label{tab:ethnicity_full}
  \begin{tabular}{lccc}
     \hline
    & \multicolumn{3}{c}{Full property rental} \\
    \hline
    AMS & W & A & B \\
    \hline
    W & [76.73; 76.75]\% & [7.62; 7.64]\% & [1.14; 1.15]\% \\
    	& 76.81\% $\color{green} \uparrow$ & 7.59\% $\color{red} \downarrow$ & 1.14\%  \\
    \hline
    A & [4.07; 4.08]\% & [0.40; 0.41]\% & [0.06; 0.06]\% \\
       & 4.07\% $\color{red} \downarrow$ & 0.43\% $\color{green} \uparrow$ & 0.06\% \\
    \hline
    B & [1.88; 1.89]\% & [0.18; 0.19]\% & [0.03; 0.03]\% \\
       & 1.87\% $\color{red} \downarrow$ & 0.20\% $\color{green} \uparrow$ & 0.03\% $\color{red} \downarrow$ \\
    \hline
    
    CHI & W & A & B \\
    \hline
    W & [71.66; 71.71]\% & [9.41; 9.45]\% & [2.06; 2.08]\% \\
    	& 72.07\% $\color{green} \uparrow$ & 9.32\% $\color{red} \downarrow$ & 2.01\% $\color{red} \downarrow$ \\
    \hline
    A & [6.73; 6.77]\% & [0.88; 0.91]\% & [0.02; 0.02]\% \\
       & 6.64\% $\color{red} \downarrow$ & 0.98\% $\color{green} \uparrow$ & 0.02\% \\
    \hline
    B & [1.71; 1.73]\% & [0.22; 0.23]\% & [0.04; 0.05]\%  \\
       & 1.50\% $\color{red} \downarrow$ & 0.24\% $\color{green} \uparrow$ & 0.15\% $\color{green} \uparrow$ \\
    \hline

    DUB & W & A & B \\
    \hline
    W & [82.34; 82.42]\% & [6.98; 7.05]\% & [1.52; 1.55]\% \\
    	& 82.28\% $\color{red} \downarrow$ & 7.10\% $\color{green} \uparrow$ & 1.43\% $\color{red} \downarrow$ \\
    \hline
    A & [3.32; 3.38]\% & [0.28; 0.33]\% & [0.09; 0.12]\% \\
       & 3.73\% $\color{green} \uparrow$ & 0.07\% $\color{red} \downarrow$ & 0.00\% $\color{red} \downarrow$ \\
    \hline
    B & [0.87; 0.90]\% & [0.10; 0.13]\% & [0.07; 0.09]\% \\
       & 0.57\% $\color{red} \downarrow$ & 0.22\% $\color{green} \uparrow$ & 0.22\% $\color{green} \uparrow$ \\
    \hline

    HK & W & A & B \\
    \hline
    W & [35.72; 35.82]\% & [22.39; 22.49]\% & [0.74; 0.76]\%  \\
    	& 37.41\% $\color{green} \uparrow$ & 20.74\% $\color{red} \downarrow$ & 0.72\% $\color{red} \downarrow$ \\
    \hline
    A & [20.05; 20.15]\% & [12.47; 12.57]\% & [0.40; 0.42]\% \\
       & 18.32\% $\color{red} \downarrow$ & 14.34\% $\color{green} \uparrow$ & 0.38\% $\color{red} \downarrow$ \\
    \hline
    B & [1.03; 1.06]\% & [0.62; 0.65]\% & [0.03; 0.03]\% \\
       & 1.12\% $\color{green} \uparrow$ & 0.57\% $\color{red} \downarrow$ & 0.05\% $\color{green} \uparrow$ \\
    \hline
 
    NAS & W & A & B \\
    \hline
    W & [78.95; 79.04]\% & [6.66; 6.73]\% & [2.14; 2.18]\%  \\
    	& 78.92\% $\color{red} \downarrow$ & 6.52\% $\color{red} \downarrow$ & 2.21\% $\color{green} \uparrow$ \\
    \hline
    A & [2.83; 2.89]\% & [0.22; 0.26]\% & [0.08; 0.10]\%  \\
       & 2.96\% $\color{green} \uparrow$ & 0.22\% & 0.00\% $\color{red} \downarrow$ \\
    \hline
    B & [3.36; 3.42]\% & [0.24; 0.29]\% & [0.10; 0.12]\%  \\
       & 3.40\%  & 0.38\% $\color{green} \uparrow$ & 0.11\%  \\
    \hline
                 
  \end{tabular}
\end{table}

\begin{table}[!h]
 \small
  \caption{Pairings between racial backgrounds of Airbnb guests and hosts (W=White, A=Asian, B=Black) in shared property rentals. Values in brackets represent 95\% confidence level intervals obtained from the rewiring analysis, while values below them denote the corresponding empirically observed frequencies. Upward green (downward red) arrows highlight over-expressed (under-expressed) values.}
  \label{tab:ethnicity_shared}
  \begin{tabular}{lcccc}
     \hline
    & \multicolumn{3}{c}{Shared property rental} \\
    \hline
    AMS  W & A & B \\
    \hline
    W & [72.34: 72.37]\% & [10.37; 10.40]\% & [1.24; 1.25]\% \\
    & 72.39\% $\color{green} \uparrow$ & 10.31\% $\color{red} \downarrow$ & 1.24\% \\
    \hline
    A & [8.30; 8.33]\% & [1.18; 1.20]\% & [0.14; 0.15]\% \\
       & 8.27\% $\color{red} \downarrow$ & 1.24\% $\color{green} \uparrow$ & 0.13\% \\
    \hline
    B & [1.46; 1.47]\% & [0.20; 0.21]\% & [0.02; 0.03]\% \\
       & 1.44\%$\color{red} \downarrow$ & 0.24\% $\color{green} \uparrow$ & 0.03\%  \\
    \hline
    
    CHI & W & A & B \\
    \hline
    W & [66.48; 66.53]\% & [12.87; 12.92]\% & [2.11; 2.13]\% \\
    	& 66.92\% $\color{green} \uparrow$ & 12.57\% $\color{red} \downarrow$ & 2.03\% $\color{red} \downarrow$ \\
    \hline
    A & [7.73; 7.74]\% & [1.49; 1.53]\% & [0.24; 0.26]\% \\
       & 7.32\% $\color{red} \downarrow$ & 1.89\% $\color{green} \uparrow$ & 0.28\% $\color{green} \uparrow$ \\
    \hline
    B & [2.04; 2.07]\% & [0.40; 0.42]\% & [0.06; 0.07]\% \\
       & 1.96\% $\color{red} \downarrow$ & 0.39\% $\color{red} \downarrow$ & 0.16\% $\color{green} \uparrow$ \\
    \hline

    DUB & W & A & B \\
    \hline
    W & [80.20; 80.29]\% & [6.60; 6.67]\% & [0.79; 0.82]\% \\
    	& 80.20\% & 6.69\% $\color{green} \uparrow$ & 0.75\% $\color{red} \downarrow$ \\
    \hline
    A & [5.11; 5.18]\% & [0.39; 0.44]\% & [0.08; 0.10]\% \\
       & 5.19\% $\color{green} \uparrow$ & 0.27\% $\color{red} \downarrow$ & 0.16\% $\color{green} \uparrow$ \\
    \hline
    B & [1.07; 1.11]\% & [0.11; 0.13]\% & [0.05; 0.05]\% \\
       & 1.07\%  & 0.11\%  & 0.00\% $\color{red} \downarrow$ \\
    \hline

    HK & W & A & B  \\
    \hline
    W & [29.04; 29.16]\% & [25.74; 25.86]\% & [0.81; 0.83]\% \\
    	& 30.72\% $\color{green} \uparrow$ & 23.90\% $\color{red} \downarrow$ & 0.84\% $\color{green} \uparrow$\\
    \hline
    A & [19.64; 19.76]\% & [17.52; 17.64]\% & [0.54; 0.57]\% \\
       & 18.04\% $\color{red} \downarrow$ & 19.51\% $\color{green} \uparrow$ & 0.52\% $\color{red} \downarrow$
       \\
    \hline
    B & [0.90; 0.92]\% & [0.77; 0.80]\% & [0.03; 0.04]\% \\
       & 0.81\% $\color{red} \downarrow$ & 0.87\% $\color{green} \uparrow$ & 0.06\% $\color{green} \uparrow$ \\
    \hline
 
    NAS & W & A & B \\
    \hline
    W & [76.54; 76.69]\% & [6.26; 6.37]\% & [2.05; 2.11]\% \\
        & 77.13\% $\color{green} \uparrow$ & 6.60\% $\color{green} \uparrow$ & 1.49\% $\color{red} \downarrow$ \\
    \hline
    A & [5.11; 5.22]\% & [0.39; 0.47]\% & [0.17; 0.22]\% \\
       & 4.79\% $\color{red} \downarrow$ & 0.21\% $\color{red} \downarrow$ & 0.43\% $\color{green} \uparrow$ \\
    \hline
    B & [1.88; 1.95]\% & [0.16; 0.21]\% & [0.11; 0.15]\% \\
       & 1.70\% $\color{red} \downarrow$ & 0.21\% & 0.32\% $\color{green} \uparrow$ \\
    \hline
                 
  \end{tabular}
\end{table}

\section{Discussion}   
In this section, we discuss the results presented above, trying to offer an interpretation based on their societal context, and proposing recommendations concerning the design of sharing economy platforms. We break down the discussion into two parts: we first consider the Airbnb's user base in each city under study, and reflect upon it relative to the city's demographic, economic, and historical context; we then move our discussion to our findings on pairing dynamics. 

\subsection{City Demographics vs. Airbnb Community}
Echoing the findings from other studies of the sharing economy, our investigation into the user base of Airbnb revealed a disparity between its communities and the city-level demographics surrounding them, both in terms of age, gender and race. As far as {\em age} is concerned, we found Airbnb hosts and guests to be overwhelmingly young (mid-twenties to mid-thirties). This can be interpreted as a reflection of the broader age-related digital divide phenomenon \cite{white2013moving}.

In terms of {\em gender}, we have found the Airbnb community to be predominantly female. In 2015, Airbnb reported that 54\% \cite{airbnb2015} of their guests were female. Based on the data we have collected up to 2017 for the cities under study, such percentage seems to be substantially higher (60\% and above), both for hosts and for guests, both for private and for shared rental properties. We do not know whether this is a signal of an evolutionary mechanism, whereby more female than male users join the platform attracted by homophily, as they already see more female users already being engaged with it. 

Perhaps the most notable results were found in terms of {\em race}. In the following, we focus our attention only on hosts, so to compare our results with available census information on the demographics of the cities under study. In all cities, the majority of hosts were found to be racially White. In particular, Dublin was found to have the highest proportion of White hosts (at 96\% for full property rental); this is expected for a city whose resident population was reported to be 90\% racially White in the latest Census\footnote{{http://www.cso.ie/en/census/census2011reports/census2011thisisirelandpart1/}} (2011).

Things are considerably different in the more diverse cities we analyzed, where we found well-known inequalities along racial lines to be largely replicated in Airbnb's interactions, with systematic evidence of a divide between the Airbnb community and the local demographics. Indeed, we found up to 64\% of Airbnb hosts in Hong Kong to be White (for full property rentals), even when 92\% of the population is reported to be of Chinese (i.e., Asian) racial background as per the 2016 Census\footnote{{http://www.bycensus2016.gov.hk/en/bc-mt.html}}. Hong Kong is well-known to be plagued by rising wealth inequality \cite{hk13}  and exorbitant property prices, that are known to be among the most unaffordable in major cities \cite{fthk}. This is coupled with high income inequality, with Hong Kong's racially White population amassing relatively high household income, while the same does not hold for the majority of the local Chinese population \cite{hk14}. Owning a spare property to rent out might thus be a privilege mostly in the hands of the White population.

Similarly of interest is Amsterdam's large majority of racially White hosts, resting at roughly 90\% of the total Airbnb user pool. This too is substantially higher than expected, given that a third of Amsterdam's population is composed of migrants recognized to be of non-western racial origins\footnote{{https://www.cbs.nl/nl-nl/achtergrond/2016/47/bevolking-naar-migratieachtergrond}}. This finding is in line with research conducted by the Netherlands' Central Commission for Statistics (CCS), which has highlighted the existence of societal integration issues \cite{ams14} amongst the Netherland's non-western population, which we speculate to be reflected into a decreased ability to secure properties to rent out.

Chicago and Nashville were found to have the highest proportion of Black hosts and guests recorded among the cities under study, averaging around 3-4\% across the two property rental types. However, this result too is notable, given that the most recent census reports Chicago's and Nashville's total populations to be 32\%\footnote{{https://www.infoplease.com/node/70107}} and 28\%\footnote{https://suburbanstats.org/population/tennessee/how-many-people-live-in-nashville} racially Black, respectively -- an order of magnitude more than what found on Airbnb. This resonates with recent research suggesting that Airbnb is a conduit for racial gentrification where the old, local community members of a neighborhood lose out in housing and in wealth \cite{gentri}.

\subsection{Homophily and Avoidance}

Our statistical investigation on pairing dynamics detected evidence of homophily both in terms of gender and race. Gender homophily is well documented to be `built in' even in young children \cite{shrum1988friendship,stehle2013gender}, so it is not surprising we could detect it in our results too. Conversely, it was interesting to see it supplanted by heterophilous behavior (i.e., a statistical over-expression of interactions between hosts and guests of different genders) in the case of Nashville, regardless of the property type, and both in Dublin and Hong Kong when switching from full property to shared property rentals. This is even more interesting when considering that full properties obviously do not imply any shared space between hosts and guests, and often allow to avoid any live interaction through automated checkin procedures. We speculate that a possible explanation behind this might lie in the different communities that naturally self-select based on property type, with those selecting shared accommodation most likely being more open-minded and prone to meeting different people (see \cite{klein2017quality} for similar findings in couchsurfing platforms).

In addition, we also detected less strong statistical signals of homophily when analysing pairing dynamics based on racial background. Once again, this is somewhat to be expected, as racial homophily has been detected in a broad variety of social environments \cite{mcpherson2001birds}, ranging from labour markets \cite{jacquemet2012indiscriminate} to online social networks \cite{wimmer2010beyond}. Symmetrical to this, we also detected phenomena of racial avoidance (still with quite weak statistical signals in most cases), i.e., under-expressions of relationships between guests and hosts belonging to different racial groups. This, again, resonates with pre-existing literature. For example, racial avoidance has been found to partially explain relocation patterns within countries \cite{rathelot2014local}. These results were accompanied by a few exceptions where we detected under-expressed homophily (White-to-White relationships in Dublin's and Nashville's full properties) and heterophily (e.g., Asian-to-White relationships in Dublin).

Since our results are of a purely statistical nature, we can only highlight what relationships are over/under-represented, without making any claims on causality. In particular, we are in no position to distinguish between avoidance and outright discrimination. Yet, some of the trends we observe are worrying and raise questions about potential countermeasures that platforms might adopt in order to monitor their progress and possibly control them. For example, following research on unconscious biases \cite{lai2014reducing}, platforms could design interventions aimed at providing users with detailed information about the peers they chose to interact with (or not) in the past, possibly highlighting systematic preferences or deviations from the outcomes that would be obtained under an unbiased selection process. Interventions could also go a step beyond raising awareness of individual behaviours. For example, they could encourage behaviours to enhance heterophily (which we already detected in some cities) by means of incentive systems similar to those that are already in place to promote service excellence (e.g., rewarding outstanding Airbnb hosts with a `superhost' status). In this fashion, users with a history of interactions with peers from different racial backgrounds could be rewarded with badges or statuses highlighting their role as diversity champions. Last but not least, platforms could incentivise users to give up potentially unnecessary steps in the interaction process where additional, and potentially biasing, information about other peers is usually acquired; for example, Airbnb's `instant booking' option, where a guest's request is automatically accepted by the platform, without an explicit consent action from the host, has been an exemplary step in this direction.

\subsection{Limitations}          

We ought to acknowledge three main limitations of this work: first, in the generation of the null network models, we have not enforced the preservation of temporal constraints (i.e., it is possible for a stay that occurred between guest $g_1$ and host $h_1$ in year $y_1$ to be swapped with a stay between $g_1$ and $h_2$, despite host $h_2$ joining Airbnb only in year $y_2 > y_1$). We chose to adopt this simplified approach under the assumption that Airbnb demographics composition has not changed significantly between 2008 and 2016 (e.g., women have consistently made up the majority of the Airbnb host community \cite{airbnb2017women}). In the future, we will consider generating null network models that preserve timing constraints (see, e.g., \cite{li2018reciprocity}).

Second, our findings rely on the accuracy of several image processing tools, to automatically annotate profile pictures in terms of gender, age and race. If the accuracy of these annotations is low, then the findings are void. In this paper, we have tried to reduce this risk by cross validating annotations across several image processing tools, and by verifying the robustness of our findings with respect to variations in their accuracy. Even so, we had to disregard any user whose profile picture did not present a (recognisable) human face, or where the estimated confidence of the annotation was low. Platform owners are most likely in possession of more accurate demographic information, explicitly provided at the time of user registration; they could thus skip the image annotation step (Airbnb Data section) and directly use this information to annotate nodes in the bipartite graph, then proceeding with the application of the statistical network analysis method we proposed (Method section) to extract more robust results.

Finally, we ought to acknowledge that our analysis of Airbnb pairing dynamics was limited to what the platform makes externally visible (i.e., reviews that hosts and guests leave to one another after a stay); results might differ if one had the opportunity to apply our method to the whole history of interactions (including stays that resulted in no reviews, and reservation requests that were cancelled/refused). Once again, platform owners do possess the whole interaction history, and might thus want to repeat this study so to validate our findings on a complete network of host/guest stays. Yet, we have reason to believe our results would hold regardless. Indeed, the large samples our analysis relies on are such that only major differences in the tendency to leave reviews between groups would affect the significance of the findings reported in this paper.      
                   
\section{Conclusion}
In this paper, we have gathered and analyzed data to assess the diversity of the Airbnb community, especially with respect to the cities where it is embedded, and we have presented a method based on the statistical analysis of networks to detect homophily, heterophily and avoidance between different groups in the Airbnb community. To the best of our knowledge, network rewiring techniques, and, more generally, null network ensembles have never been employed as a tool to detect bias, and this application represents an element of novelty of our work.

Our findings suggest that, in all cities under study, certain user groups (e.g., young, White, female) are substantially over-represented compared to the local population; furthermore, statistically significant signals of gender and racial homophily were detected, across all cities and regardless of the property rental type.

Taken together, our findings echo sentiment that perhaps, contrary to all branding, the sharing economy community might not be that diverse. Rather, platforms such as Airbnb might be acting as accelerators for gentrification processes that are already well underway in major cities. While policy and legislation interventions are needed to regulate who benefits from sharing economy platforms \cite{quattrone2016benefits}, technological considerations also deserve attention: for example, Airbnb, like most sharing economy platforms, requires hosts to have a traditional bank account at the time of registration, to which money will then be deposited when guests visit. This might hinder the ability for many hosts from socio-economic deprived backgrounds to join the platform in the first place (which might be reflected in the very low representation from certain racial backgrounds in our studies); alternative solutions, such as on-demand payment services provided by platforms like BitPesa,\footnote{{http://www.bitpesa.co/}} could lower the barrier to entry.

\begin{acknowledgements}
W.L. and G.L. were supported by an EPSRC Early Career Fellowship in Digital Economy (Grant No. EP/N006062/1).
\end{acknowledgements}

\bibliography{KLLC_refs} 

\appendix

\section{Robustness with respect to annotation accuracy}
\label{app_robust}

The results presented in the main paper have been obtained by using two facial annotation softwares (Sightcorp and Betaface) in parallel. The user pictures retained in our dataset were only those for which both products provided the same annotations with a confidence higher than $0.3 \in [0,1]$. As a robustness test of our results, we repeated our analyses on a restricted dataset limited to images for which both softwares provided the same annotations with a confidence higher than $0.5$. 

The summary statistics about the number of hosts, guests and pairings that we collected and annotated with confidence $\geq 0.5$ for each city are reported in Table \ref{tab:annotation}. The number of annotated users decreases slightly with the higher confidence threshold, but the demographic features of this subset of data remain largely unchanged with respect to those reported in the main paper (see Tables \ref{tab:features_gender} and \ref{tab:features_ethnicity}).

In Tables \ref{tab:gender} and \ref{tab:ethnicity} we report the results obtained from the rewiring analysis on this restricted dataset on gender and race, respectively. As it can be seen, the over- and under-expression patterns we find in the gender-related pairings are exactly the same as those obtained with a lower accuracy threshold reported in the main paper. The same applies to the race-related pairings, which we find to be consistent with those obtained with a lower threshold in all but a few cases (see, e.g., Black-White pairings in Nasvhille's full property rentals).

\begin{table}[!h]
  \caption{Number of hosts, guests, and host-guest pairs annotated for each city analysed when setting the annotation confidence threshold to 0.5}
  \centering
  \label{tab:annotation}
  \begin{tabular}{lccc}
    \hline
    City & \# Hosts & \# Guests & \# Host-Guest Pairs \\
    \hline
    Amsterdam & 2,349 &  68,978   &  70,812  \\
    Chicago & 1,685  & 20,719  & 22,077  \\
    Dublin & 1,033  & 2,590  & 2,757  \\
    Hong Kong & 1,222  & 11,915  & 13,077  \\
    Nashville & 626  & 1,702  & 2,000  \\
    \hline
  \end{tabular}
\end{table}

\begin{table}[!h]
  \caption{Airbnb host and guest population by gender (F=female) in the dataset restricted to images annotated with confidence higher than 0.5}
  \centering
  \label{tab:features_gender}
  \begin{tabular}{lcc|cc}
    \hline
    & \multicolumn{2}{c|}{Full property rental} & \multicolumn{2}{c}{Shared property rental} \\
    \hline
    City & F Host & F Guest & F Host & F Guest \\
   \hline
    Amsterdam & 59\% & 55\% & 58\% & 58\%  \\
    Chicago   & 62\% & 58\% & 57\% & 56\%  \\
    Dublin    & 60\% & 59\% & 59\% & 54\%  \\
    Hong Kong & 55\% & 59\% & 60\% & 59\%  \\
    Nashville & 63\% & 66\% & 74\% & 59\%  \\
    \hline
  \end{tabular}
  %}
\end{table}

\begin{table}[!h]
  \caption{Airbnb host population by race (W=White, A=Asian, B=Black) in the dataset restricted to images annotated with confidence higher than 0.5}
  \centering
  \label{tab:features_ethnicity}
%  \resizebox{\columnwidth}{!}{
  \begin{tabular}{lccc|ccc}
    \hline
    & \multicolumn{3}{c|}{Full property rental} & \multicolumn{3}{c}{Shared property rental} \\
    \hline
    City & W & A & B & W & A & B \\
    \hline
    Amsterdam &  94\% & 4\%  & 2\% &  90\% & 9\% & 1\% \\
    Chicago   &  91\% & 7\%  & 2\% &  89\% & 9\% & 2\% \\
    Dublin    &  96\% & 3\%  & 1\% &  93\% & 5\% & 2\% \\
    Hong Kong &  65\% & 34\% & 1\% &  59\% & 40\% & 1\% \\
    Nashville &  93\% & 3\%  & 4\% &  92\% &  5\% & 3\% \\
    \hline
  \end{tabular}
  %}
\end{table}

\begin{small}
\begin{table}[h!]

  \caption{Pairings between guest-host genders (F=female; M=male) in the dataset restricted to images annotated with confidence higher than 0.5. Values in brackets represent 95\% confidence level intervals obtained from the rewiring analysis, while the values to their right denote the corresponding empirically observed frequencies. Upward green (downward red) arrows highlight over-expressed (under-expressed) values.}
  \label{tab:gender}

   \hspace{-2cm} \begin{tabular}{lclclclcl}
    \hline
    
    & \multicolumn{8}{c}{Full property rental} \\
    \hline
    City & FF & FF & FM & FM & MF & MF & MM & MM  \\
    \hline
    AMS & [29.40; 29.45]\% &  30.23\% $\color{green} \uparrow$ & [24.01; 24.05]\%  & 23.23\% $\color{red} \downarrow$ & [25.60; 25.64]\% & 24.81\% $\color{red} \downarrow$ & [20.91; 20.95]\% & 21.73\% $\color{green} \uparrow$ \\
    CHI & [33.71; 33.79]\% & 34.90\% $\color{green} \uparrow$ & [25.07; 25.15]\% & 23.96\% $\color{red} \downarrow$ & [23.58; 23.66]\% & 22.47\% $\color{red} \downarrow$ & [17.47; 17.55]\% & 18.67\% $\color{green} \uparrow$ \\
    DUB & [34.85; 35.11]\% & 37.28\% $\color{green} \uparrow$ & [25.14; 25.40]\% & 22.97\% $\color{red} \downarrow$ & [22.82; 23.07]\% & 20.64\% $\color{red} \downarrow$ & [16.68; 16.94]\% & 19.11\% $\color{green} \uparrow$ \\
    HK & [33.18; 33.29]\% & 33.79\% $\color{green} \uparrow$ & [24.40; 24.51]\% & 23.91\% $\color{red} \downarrow$ & [24.32; 24.43]\% & 23.82\% $\color{red} \downarrow$ & [17.88; 17.99]\% & 18.49\% $\color{green} \uparrow$ \\
    NAS & [41.46; 41.67]\% & 41.94\% $\color{green} \uparrow$ & [22.58; 22.79]\% & 22.30\% $\color{red} \downarrow$ & [23.06; 23.28]\% & 22.79\% $\color{red} \downarrow$ & [12.48; 12.69]\% & 12.97\% $\color{green} \uparrow$ \\
    \hline
    & \multicolumn{8}{c}{Shared property rental} \\  
    \hline    
        AMS & [28.52; 28.58]\% & 29.29\% $\color{green} \uparrow$ & [20.57; 20.63]\%  & 19.86\% $\color{red} \downarrow$ & [29.47; 29.53]\% & 28.76\% $\color{red} \downarrow$ & [21.32; 21.39]\% & 22.09\% $\color{green} \uparrow$ \\
        CHI & [31.36; 31.45]\% & 32.73\% $\color{green} \uparrow$ & [24.64; 24.74]\% & 23.36\% $\color{red} \downarrow$ & [24.53; 24.62]\% & 23.25\% $\color{red} \downarrow$ & [19.28; 19.38]\% & 20.66\% $\color{green} \uparrow$ \\
    	DUB & [30.98; 31.17]\% & 29.59\% $\color{red} \downarrow$ & [26.66; 26.85]\% & 28.24\% $\color{green} \uparrow$ & [22.45; 22.64]\% & 24.03\% $\color{green} \uparrow$ & [19.53; 19.72]\% & 18.14\% $\color{red} \downarrow$ \\
    	HK & [33.55; 33.68]\% & 34.22\% $\color{green} \uparrow$ & [24.12; 24.25]\% & 23.58\% $\color{red} \downarrow$ & [24.55; 24.67]\% & 24.00\% $\color{red} \downarrow$ & [17.53; 17.66]\% & 18.20\% $\color{green} \uparrow$ \\
    NAS & [42.72; 43.00]\% & 42.84\%  & [30.52; 30.80]\% & 30.68\%  & [15.24; 15.51]\% & 15.39\%  & [10.97; 11.24]\% & 11.09\%  \\
    \hline
  \end{tabular}
  \end{table}
\end{small}

\begin{table}[!p]
 \small
  \caption{Pairings between racial backgrounds of Airbnb guests and hosts (W=White, A=Asian, B=Black) in the dataset restricted to images annotated with confidence higher than 0.5. Values in brackets represent 95\% confidence level intervals obtained from the rewiring analysis, while values below them denote the corresponding empirically observed frequencies. Upward green (downward red) arrows highlight over-expressed (under-expressed) values.}
  \label{tab:ethnicity}
  \begin{tabular}{lccccccc}
     \hline
    & \multicolumn{3}{c}{Full property rental} & \vline & \multicolumn{3}{c}{Shared property rental} \\
    \hline
    AMS & W & A & B & \vline & W & A & B \\
    \hline
    W & [77.37; 77.39]\% & [7.57; 7.59]\% & [1.06; 1.07]\% & \vline & [72.93: 72.96]\% & [10.35; 10.37]\% & [1.21; 1.21]\% \\
    & 77.46\% $\color{green} \uparrow$ & 7.52\% $\color{red} \downarrow$ & 1.06\% & \vline & 73.01\% $\color{green} \uparrow$ & 10.25\% $\color{red} \downarrow$ & 1.23\% $\color{green} \uparrow$ \\
    \hline
    A & [3.84; 3.85]\% & [0.37; 0.38]\% & [0.05; 0.06]\% & \vline & [8.00; 8.02]\% & [1.13; 1.16]\% & [0.12; 0.13]\% \\
       & 3.85\%  & 0.40\% $\color{green} \uparrow$ & 0.05\% & \vline & 7.93\% $\color{red} \downarrow$ & 1.21\% $\color{green} \uparrow$ & 0.11\% $\color{red} \downarrow$ \\
    \hline
    B & [1.87; 1.87]\% & [0.18; 0.19]\% & [0.02; 0.03]\% & \vline & [1.47; 1.48]\% & [0.20; 0.21]\% & [0.02; 0.03]\% \\
       & 1.85\% $\color{red} \downarrow$  & 0.19\% $\color{green} \uparrow$ & 0.02\% $\color{red} \downarrow$ & \vline & 1.45\%$\color{red} \downarrow$ & 0.24\% $\color{green} \uparrow$ & 0.03\%  \\
    \hline
    
    CHI & W & A & B & \vline & W & A & B \\
    \hline
    W & [72.14; 72.19]\% & [9.34; 9.38]\% & [1.98; 2.00]\% & \vline & [67.50; 67.55]\% & [12.97; 13.02]\% & [2.06; 2.08]\% \\
    	& 72.54\% $\color{green} \uparrow$ & 9.23\% $\color{red} \downarrow$ & 1.93\% $\color{red} \downarrow$ & \vline & 67.90\% $\color{green} \uparrow$ & 12.70\% $\color{red} \downarrow$ & 2.00\% $\color{red} \downarrow$ \\
    \hline
    A & [6.72; 6.76]\% & [0.86; 0.89]\% & [0.02; 0.02]\% & \vline & [7.84; 7.88]\% & [1.49; 1.54]\% & [0.23; 0.25]\% \\
       & 6.62\% $\color{red} \downarrow$ & 0.96\% $\color{green} \uparrow$ & 0.02\%  & \vline & 7.46\% $\color{red} \downarrow$ & 1.89\% $\color{green} \uparrow$ & 0.27\% $\color{green} \uparrow$ \\
    \hline
    B & [1.64; 1.66]\% & [0.21; 0.22]\% & [0.04; 0.05]\% & \vline & [1.61; 1.63]\% & [0.31; 0.32]\% & [0.04; 0.05]\% \\
       & 1.45\% $\color{red} \downarrow$ & 0.23\% $\color{green} \uparrow$ & 0.13\% $\color{green} \uparrow$ & \vline & 1.55\% $\color{red} \downarrow$ & 0.30\% $\color{red} \downarrow$ & 0.12\% $\color{green} \uparrow$ \\
    \hline

    DUB & W & A & B & \vline & W & A & B \\
    \hline
    W & [83.04; 83.13]\% & [6.88; 6.95]\% & [1.56; 1.59]\% & \vline & [80.46; 80.55]\% & [6.31; 6.39]\% & [0.76; 0.78]\% \\
    	& 82.99\% $\color{red} \downarrow$ & 6.98\% $\color{green} \uparrow$ & 1.45\% $\color{red} \downarrow$ & \vline & 80.51\% & 6.37\% & 0.70\% $\color{red} \downarrow$ \\
    \hline
    A & [3.30; 3.38]\% & [0.26; 0.32]\% & [0.09; 0.11]\% & \vline & [5.18; 5.24]\% & [0.39; 0.44]\% & [0.08; 0.10]\% \\
       & 3.78\% $\color{green} \uparrow$ & 0.00\% $\color{red} \downarrow$ & 0.00\% $\color{red} \downarrow$ & \vline & 5.24\%  & 0.27\% $\color{red} \downarrow$ & 0.16\% $\color{green} \uparrow$ \\
    \hline
    B & [0.89; 0.92]\% & [0.10; 0.12]\% & [0.07; 0.08]\% & \vline & [1.10; 1.13]\% & [0.10; 0.13]\% & [0.05; 0.07]\% \\
       & 0.58\% $\color{red} \downarrow$ & 0.22\% $\color{green} \uparrow$ & 0.22\% $\color{green} \uparrow$ & \vline & 1.08\% $\color{red} \downarrow$ & 0.11 & 0.00\% $\color{red} \downarrow$ \\
    \hline

    HK & W & A & B & \vline & W & A & B \\
    \hline
    W & [35.87; 35.99]\% & [22.39; 22.50]\% & [0.69; 0.71]\% & \vline & [28.85; 28.96]\% & [25.73; 25.84]\% & [0.75; 0.77]\% \\
    	& 37.56\% $\color{green} \uparrow$ & 20.76\% $\color{red} \downarrow$ & 0.67\% $\color{red} \downarrow$ & \vline & 30.66\% $\color{green} \uparrow$ & 23.79\% $\color{red} \downarrow$ & 0.79\% $\color{green} \uparrow$\\
    \hline
    A & [20.27; 20.38]\% & [12.62; 12.73]\% & [0.39; 0.41]\% & \vline & [19.96; 20.07]\% & [17.61; 17.73]\% & [0.49; 0.52]\% \\
       & 18.54\% $\color{red} \downarrow$ & 14.52\% $\color{green} \uparrow$ & 0.37\% $\color{red} \downarrow$ & \vline & 18.21\% $\color{red} \downarrow$ & 19.70\% $\color{green} \uparrow$ & 0.46\% $\color{red} \downarrow$
       \\
    \hline
    B & [0.80; 0.82]\% & [0.48; 0.51]\% & [0.02; 0.03]\% & \vline & [0.90; 0.94]\% & [0.79; 0.82]\% & [0.03; 0.03]\% \\
       & 0.87\% $\color{green} \uparrow$ & 0.43\% $\color{red} \downarrow$ & 0.04\% $\color{green} \uparrow$ & \vline & 0.82\% $\color{red} \downarrow$ & 0.88\% $\color{green} \uparrow$ & 0.06\% $\color{green} \uparrow$ \\
    \hline
 
    NAS & W & A & B & \vline & W & A & B \\
    \hline
    W & [79.27; 79.36]\% & [6.63; 6.70]\% & [2.04; 2.08]\% & \vline & [77.23; 77.37]\% & [6.29; 6.40]\% & [2.13; 2.18]\% \\
    	& 79.16\% $\color{red} \downarrow$ & 6.57\% $\color{red} \downarrow$ & 2.12\% $\color{green} \uparrow$ & \vline & 77.83\% $\color{green} \uparrow$ & 6.67\% $\color{green} \uparrow$ & 1.51\% $\color{red} \downarrow$ \\
    \hline
    A & [2.85; 2.90]\% & [0.23; 0.28]\% & [0.09; 0.11]\% & \vline & [5.23; 5.35]\% & [0.41; 0.50]\% & [0.15; 0.18]\% \\
       & 2.98\% $\color{green} \uparrow$ & 0.22\% $\color{red} \downarrow$ & 0.00\% $\color{red} \downarrow$ & \vline & 4.84\% $\color{red} \downarrow$ & 0.22\% $\color{red} \downarrow$ & 0.43\% $\color{green} \uparrow$ \\
    \hline
    B & [3.28; 3.34]\% & [0.25; 0.30]\% & [0.10; 0.12]\% & \vline & [1.92; 1.99]\% & [0.17; 0.21]\% & [0.11; 0.15]\% \\
       & 3.36\% $\color{green} \uparrow$ & 0.32\% $\color{green} \uparrow$ & 0.11\% & \vline & 1.72\% $\color{red} \downarrow$ & 0.22\% $\color{green} \uparrow$  & 0.32\% $\color{green} \uparrow$ \\
    \hline
                 
  \end{tabular}
\end{table}

As an additional robustness check of our results, we repeated our analysis on race-related pairings after an artificial manipulation of the data. Namely, we manually altered the race annotation of a randomly selected sample made of $5\%$ of all White users in each city, changing their annotation to Black or Asian with probability $1/2$. The results obtained from the rewiring analysis on this manipulated dataset are shown in Table \ref{tab:ethnicity_mani}. As one might expect, this leads to a few changes with respect to the results presented in the main paper. For example, the over-expression of White-White interactions is  eliminated in the case of Amsterdam. However, most of the homophily, heterophily, and avoidance patterns reported in the main paper do not change.
  
%We randomly select $2.5\%$ white users to be black and $2.5\%$ white users to be asian. Then we do the rewiring analysis to see how error-tolerant the main findings of our research is.

\begin{table}[!p]
 \small
  \caption{Pairings between racial backgrounds of Airbnb guests and hosts (W=White, A=Asian, B=Black) in a manipulated dataset where a randomly selected sample of $5\%$ of all White users are artificially annotated as either Black or Asian with probability $1/2$. Values in brackets represent 95\% confidence level intervals obtained from the rewiring analysis, while values below them denote the corresponding empirically observed frequencies. Upward green (downward red) arrows highlight over-expressed (under-expressed) values.}
  \label{tab:ethnicity_mani}
  \begin{tabular}{lccccccc}
     \hline
    & \multicolumn{3}{c}{Full property rental} & \vline & \multicolumn{3}{c}{Shared property rental} \\
    \hline
    AMS & W & A & B & \vline & W & A & B \\
    \hline
    W & [70.08; 70.11]\% & [9.12; 9.14] \% & [2.97; 2.98]\% & \vline & [66.31; 66.35]\% & [11.72; 11.75]\% & [2.91; 2.93]\% \\
    & 70.10\%  & 9.11 \% $\color{red} \downarrow$ & 1.03\% $\color{red} \downarrow$ & \vline & 66.27\% $\color{red} \downarrow$ & 11.72 \% $\color{red} \downarrow$ & 2.95\% \\
    \hline
    A & [5.00; 5.01]\% & [0.64; 0.66]\% & [0.21; 0.22]\% & \vline & [9.64; 9.67]\% & [1.70; 1.73]\% & [0.42; 0.44]\% \\
       & 5.04\%  $\color{green} \uparrow$ & 0.66\% $\color{green} \uparrow$ & 0.20\% $\color{red} \downarrow$ & \vline & 9.70\% $\color{green} \uparrow$ & 1.73\% $\color{green} \uparrow$ & 0.36\% $\color{red} \downarrow$ \\
    \hline
    B & [3.48; 3.49]\% & [0.45; 0.46]\% & [0.14; 0.15]\% & \vline & [2.08; 2.10]\% & [0.36; 0.37]\% & [0.09; 0.09]\% \\
       & 3.49\%   & 0.13\% $\color{red} \downarrow$ & 0.45\% $\color{green} \uparrow$ & \vline & 2.08\% & 0.38\% $\color{green} \uparrow$ & 0.10\% $\color{green} \uparrow$ \\
    \hline
    
    CHI & W & A & B & \vline & W & A & B \\
    \hline
    W & [65.03; 65.09]\% & [10.46; 10.51]\% & [3.85; 3.89]\% & \vline & [59.30; 59.36]\% & [13.48; 13.54]\% & [3.53; 3.55]\% \\
    	& 65.41\% $\color{green} \uparrow$ & 10.41\% $\color{red} \downarrow$ & 3.85\%  & \vline & 59.60\% $\color{green} \uparrow$ & 13.26\% $\color{red} \downarrow$ & 3.50\% $\color{red} \downarrow$ \\
    \hline
    A & [7.77; 7.81]\% & [1.25; 1.28]\% & [0.45; 0.48]\% & \vline & [9.51; 9.56]\% & [2.14; 2.18]\% & [0.56; 0.58]\% \\
       & 7.71 \% $\color{red} \downarrow$ & 1.32\% $\color{green} \uparrow$ & 0.42\% $\color{red} \downarrow$ & \vline & 9.16\% $\color{red} \downarrow$ & 2.55 \% $\color{green} \uparrow$ & 0.53\% $\color{red} \downarrow$ \\
    \hline
    B & [3.34; 3.37]\% & [0.53; 0.55]\% & [0.19; 0.20]\% & \vline & [3.79; 3.82]\% & [0.87; 0.90]\% & [0.22; 0.24]\% \\
       & 3.11\% $\color{red} \downarrow$ & 0.56\% $\color{green} \uparrow$ & 0.29\% $\color{green} \uparrow$ & \vline & 3.79\% $\color{red} \downarrow$ & 0.81\% $\color{red} \downarrow$ & 0.10\% $\color{red} \downarrow$ \\
    \hline

    DUB & W & A & B & \vline & W & A & B \\
    \hline
    W & [74.05; 74.17]\% & [8.72; 8.83]\% & [2.79; 2.85]\% & \vline & [72.36; 72.48]\% & [8.34; 8.43]\% & [3.14; 3.20]\% \\
    	& 74.03 \% $\color{red} \downarrow$ & 8.82\% & 2.80 \% $\color{red} \downarrow$ & \vline & 72.23 \% $\color{red} \downarrow$ & 8.56\% $\color{green} \uparrow$ & 3.10 \%  $\color{red} \downarrow$ \\
    \hline
    A & [5.43; 5.53]\% & [0.61; 0.70]\% & [0.20; 0.26]\% & \vline & [6.81; 6.90]\% & [0.76; 0.83]\% & [0.30; 0.35]\% \\
       & 5.88 \% $\color{green} \uparrow$ & 0.50\% $\color{red} \downarrow$ & 0.14\% $\color{red} \downarrow$ & \vline & 7.17\% $\color{green} \uparrow$ & 0.54\% $\color{red} \downarrow$ & 0.32\%  \\
    \hline
    B & [3.03; 3.10]\% & [0.33; 0.39]\% & [0.13; 0.17]\% & \vline & [2.19; 2.24]\% & [0.24; 0.29]\% & [0.09; 0.11]\% \\
       & 2.65\% $\color{red} \downarrow$  & 0.50 \% $\color{green} \uparrow$  & 0.29 \% $\color{green} \uparrow$  & \vline & 2.14 \% $\color{red} \downarrow$ & 0.27\% & 0.11\% \\
    \hline

    HK & W & A & B & \vline & W & A & B \\
    \hline
    W & [31.78; 31.89]\% & [22.03; 22.14]\% & [1.52; 1.55]\% & \vline & [25.97; 26.09]\% & [24.82; 24.94]\% & [1.40; 1.43]\% \\
    	& 33.22 \% $\color{green} \uparrow$ & 20.64\% $\color{red} \downarrow$ & 1.54\%  & \vline & 27.38 \% $\color{green} \uparrow$ & 23.15 \% $\color{red} \downarrow$ & 1.49 \% $\color{green} \uparrow$\\
    \hline
    A & [19.59; 19.70]\% & [13.48; 13.59]\% & [0.93; 0.96]\% & \vline & [20.06; 20.18]\% & [19.24; 19.37]\% & [1.07; 1.10]\% \\
       & 17.99 \% $\color{red} \downarrow$ & 15.17\% $\color{green} \uparrow$ & 0.94\% & \vline & 18.82\% $\color{red} \downarrow$ & 20.96\% $\color{green} \uparrow$ & 0.97\% $\color{red} \downarrow$
       \\
    \hline
    B & [2.35; 2.40]\% & [1.60; 1.65]\% & [0.11; 0.12]\% & \vline & [1.21; 1.24]\% & [1.13; 1.17]\% & [0.07; 0.08]\% \\
       & 2.54 \% $\color{green} \uparrow$ & 1.48\% $\color{red} \downarrow$ & 0.12\% $\color{green} \uparrow$ & \vline & 1.05\% $\color{red} \downarrow$ & 1.32\% $\color{green} \uparrow$  & 0.12\% $\color{green} \uparrow$  \\
    \hline
 
    NAS & W & A & B & \vline & W & A & B \\
    \hline
    W & [68.30; 68.42]\% & [8.29; 8.39]\% & [5.02; 5.10]\% & \vline & [68.77; 68.95]\% & [7.57; 7.71]\% & [3.65; 3.75]\% \\
    	& 68.09 \% $\color{red} \downarrow$ & 8.46\% $\color{green} \uparrow$ & 4.91\%  $\color{red} \downarrow$  & \vline & 68.40\% $\color{red} \downarrow$ & 8.40\% $\color{green} \uparrow$ & 3.30 \% $\color{red} \downarrow$ \\
    \hline
    A & [4.62; 4.72]\% & [0.59; 0.52]\% & [0.32; 0.37]\% & \vline & [6.77; 6.91]\% & [0.70; 0.82]\% & [0.33; 0.39]\% \\
       & 4.91 \% $\color{green} \uparrow$ & 0.49\% $\color{red} \downarrow$ & 0.27\% $\color{red} \downarrow$ & \vline & 7.13\% $\color{green} \uparrow$ & 0.21\% $\color{red} \downarrow$ & 0.43\% $\color{green} \uparrow$ \\
    \hline
    B & [6.20; 6.29]\% & [0.70; 0.77]\% & [0.43; 0.49]\% & \vline & [3.96; 4.07]\% & [0.42; 0.50]\% & [0.23; 0.28]\% \\
       & 6.31 \% $\color{green} \uparrow$ & 0.65\% $\color{red} \downarrow$ & 0.65\% $\color{green} \uparrow$ & \vline & 4.04 \%  & 0.43\%   & 0.53\% $\color{green} \uparrow$  \\
    \hline
                 
  \end{tabular}
\end{table}

\newpage

%%%%%%%%%%%%%%%%%%%%%%%%%%%%%%%%%%%%%%%%%%%%%%%%%%%%%%%%%%%%%%%%%%%%%%%           
\section{Robustness with respect to economic factors}        
\label{app_econ}   
%%%%%%%%%%%%%%%%%%%%%%%%%%%%%%%%%%%%%%%%%%%%%%%%%%%%%%%%%%%%%%%%%%%%%%%           

As mentioned in the main paper, some of the over-/under-expressions we observe in the race-related pairings might be attributed to potential confounding factors such as  users' wealth or income. Indeed, global economic inequalities often correlate with race. In particular, a White racial background typically correlates with better economic conditions. This, in turn, might partially explain some of the homophily and avoidance patterns we report in the main paper. For example, the over-expression of interactions between White hosts and guests we measured in most of the cities we analyzed could be partially (or completely) explained simply in terms of White users being on average wealthier (i.e., White hosts owning more expensive properties and White guests being able to afford more expensive stays).

We first sought to disentangle wealth and homophily via matched pair analysis. In each city we looked for pairs of White and non-White hosts with similar profiles in terms of two proxies for wealth and/or income, i.e., the number of Airbnb properties owned, and the price charged for a week-long stay at such properties. After forming such pairs (using the algorithm provided by Ref. [33] in the main paper), we measure the rate of interaction with White guests across the two groups, and run a $t$-test on the two rates against a null hypothesis of equal rates of interaction.  

The results of this analysis are reported in Table \ref{tab:mpa}. As it can be seen, we  found statistically significant differences in the rates of interaction with White guests across the two groups only in the case of Hong Kong (regardless of the type of property) and Chicago's full property rentals. This is due to the imbalance in the data, since in all cities we analyze White hosts are the vast majority, which makes it rather hard to find large enough numbers of pairs with non-White hosts. As a matter of fact, the most significant results were found in Hong Kong, which is by far the most diverse city among the ones we analyzed. 

\begin{table}[h!]
\caption{Results of the matched pair analysis}
\centering
  \label{tab:mpa}
\begin{tabular}{r|c|c|c|c|c|c}
\hline
City & Property & White/White & non-White/White & Pairs & Stays & $p$-value \\
\hline
Amsterdam & Shared & 82.31\% & 83.29\% & 56 & 2301 & 0.287\\
\hline
Amsterdam & Full & 84.88\% & 85.02\% & 187 & 4464 & 0.833\\
\hline
Chicago & Shared & 74.42\% & 75.64\% & 141 & 2905 & 0.303\\
\hline
Chicago & Full & 81.92\% & 79.45\% & 148 & 1792 & 0.0451$^{*}$\\
\hline
Dublin & Shared & 89.08\% & 90.87\% & 69 & 403 & 0.475\\
\hline
Dublin & Full & 91.89\% & 88.80\% & 33 & 111 & 0.422\\
\hline
Hong Kong & Shared & 56.21\% & 47.67\% & 217 & 3535 & $< 0.001^{***}$\\
\hline
Hong Kong & Full & 60.31\% & 53.00\% & 260 & 3051 & $< 0.001^{***}$\\
\hline
Nashville & Shared & 84.03\% & 82.08\% & 26 & 119 & 0.698\\
\hline
Nashville & Full & 85.28\% & 88.64\% & 51 & 598 & 0.233\\
\hline
\end{tabular}
\end{table}

In order to overcome this issue, we sought to control for the hosts' wealth / income by removing from each network the hosts belonging to the top and bottom third of the distribution of prices charged for a week-long stay. This left us with bipartite sub-networks made exclusively of owners of middle-range properties and their guests. We report the results of our network rewiring analysis obtained on such sub-networks in Table \ref{tab:ethnicity_sub}. Once again, we find the results to be very much in line with those reported in the main paper, with a few exceptions (most notably, interactions involving White guests in Dublin).

\begin{table}[!h]
 \small
  \caption{Pairings between racial backgrounds of Airbnb guests and hosts (W=White, A=Asian, B=Black) in the sub-networks obtained by removing all hosts belonging to the top and bottom thirds of the distribution of prices charged for a week-long stay. Values in brackets represent 95\% confidence level intervals obtained from the rewiring analysis, while values below them denote the corresponding empirically observed frequencies. Upward green (downward red) arrows highlight over-expressed (under-expressed) values.}
  \label{tab:ethnicity_sub}
  \begin{tabular}{lccccccc}
     \hline
    & \multicolumn{3}{c}{Full property rental} & \vline & \multicolumn{3}{c}{Shared property rental} \\
    \hline
    AMS & W & A & B & \vline & W & A & B \\
    \hline
    W & [80.30; 80.33]\% & [8.14; 8.16]\% & [1.02; 1.03]\% & \vline & [68.30; 68.37]\% & [7.74; 7.80]\% & [1.10; 1.13]\% \\
    & 80.33\% $\color{green} \uparrow$ & 8.09\% $\color{red} \downarrow$ & 1.03\% & \vline & 68.75\% $\color{green} \uparrow$ & 7.69\% $\color{red} \downarrow$ & 1.06\% $\color{red} \downarrow$ \\
    \hline
    A & [4.37; 4.40]\% & [0.43; 0.45]\% & [0.05; 0.06]\% & \vline & [2.64; 2.67]\% & [0.29; 0.31]\% & [0.04; 0.05]\% \\
       & 4.34\%  $\color{red} \downarrow$ & 0.52\% $\color{green} \uparrow$ & 0.06\% & \vline & 2.49\% $\color{red} \downarrow$ & 0.42\% $\color{green} \uparrow$ & 0.03\% $\color{red} \downarrow$ \\
    \hline
    B & [1.34; 1.35]\% & [0.13; 0.14]\% & [0.02; 0.02]\% & \vline & [0.85; 0.87]\% & [0.10; 0.12]\% & [0.02; 0.03]\% \\
       & 1.38\% $\color{green} \uparrow$  & 0.13\%  & 0.00\% $\color{red} \downarrow$ & \vline & 0.84\%$\color{red} \downarrow$ & 0.13\% $\color{green} \uparrow$ & 0.01\% $\color{red} \downarrow$ \\
    \hline
    
    CHI & W & A & B & \vline & W & A & B \\
    \hline
    W & [74.41; 74.48]\% & [9.59; 9.65]\% & [2.14; 2.17]\% & \vline & [69.20; 69.30]\% & [11.16; 11.25]\% & [1.59; 1.62]\% \\
    	& 74.91\% $\color{green} \uparrow$ & 9.56\% $\color{red} \downarrow$ & 2.04\% $\color{red} \downarrow$ & \vline & 69.49\% $\color{green} \uparrow$ & 11.07\% $\color{red} \downarrow$ & 1.46\% $\color{red} \downarrow$ \\
    \hline
    A & [4.57; 4.62]\% & [0.56; 0.61]\% & [0.01; 0.01]\% & \vline & [7.70; 7.78]\% & [1.25; 1.32]\% & [0.17; 0.19]\% \\
       & 4.44\% $\color{red} \downarrow$ & 0.77\% $\color{green} \uparrow$ & 0.01\%  & \vline & 7.32\% $\color{red} \downarrow$ & 1.60\% $\color{green} \uparrow$ & 0.31\% $\color{green} \uparrow$ \\
    \hline
    B & [2.24; 2.28]\% & [0.27; 0.30]\% & [0.06; 0.07]\% & \vline & [1.90; 1.93]\% & [0.29; 0.32]\% & [0.05; 0.07]\% \\
       & 1.98\% $\color{red} \downarrow$ & 0.19\% $\color{red} \downarrow$ & 0.23\% $\color{green} \uparrow$ & \vline & 1.87\% $\color{red} \downarrow$ & 0.27\% $\color{red} \downarrow$ & 0.10\% $\color{green} \uparrow$ \\
    \hline

    DUB & W & A & B & \vline & W & A & B \\
    \hline
    W & [83.15; 83.30]\% & [5.28; 5.40]\% & [1.89; 1.97]\% & \vline & [81.37; 81.52]\% & [6.44; 6.55]\% & [0.84; 0.87]\% \\
    	& 83.29\% & 5.22\% $\color{red} \downarrow$ & 2.09 \% $\color{green} \uparrow$ & \vline & 81.45\% & 6.18\% $\color{red} \downarrow$ & 0.93\% $\color{green} \uparrow$ \\
    \hline
    A & [3.96; 4.07]\% & [0.33; 0.41]\% & [0.25; 0.29]\% & \vline & [2.85; 2.93]\% & [0.26; 0.33]\% & [0.15; 0.15]\% \\
       & 4.18 \% $\color{green} \uparrow$ & 0.26\% $\color{red} \downarrow$ & 0.00\% $\color{red} \downarrow$ & \vline & 3.09\% $\color{green} \uparrow$ & 0.15\% $\color{red} \downarrow$ & 0.00\% $\color{red} \downarrow$ \\
    \hline
    B & [0.00; 0.00]\% & [0.00; 0.00]\% & [0.00; 0.00]\% & \vline & [0.54; 0.57]\% & [0.15; 0.18]\% & [0.00; 0.00]\% \\
       & 0.00\%  & 0.00 \%  & 0.00 \%  & \vline & 0.46\% $\color{red} \downarrow$ & 0.15 & 0.00\% \\
    \hline

    HK & W & A & B & \vline & W & A & B \\
    \hline
    W & [38.27; 38.42]\% & [21.64; 21.78]\% & [0.85; 0.90]\% & \vline & [28.40; 28.59]\% & [23.20; 23.38]\% & [0.74; 0.78]\% \\
    	& 39.35\% $\color{green} \uparrow$ & 20.22\% $\color{red} \downarrow$ & 0.98\% $\color{green} \uparrow$ & \vline & 29.56\% $\color{green} \uparrow$ & 21.77\% $\color{red} \downarrow$ & 0.91\% $\color{green} \uparrow$\\
    \hline
    A & [19.55; 19.70]\% & [11.12; 11.25]\% & [0.45; 0.49]\% & \vline & [22.61; 22.81]\% & [18.31; 18.50]\% & [0.58; 0.63]\% \\
       & 18.49 \% $\color{red} \downarrow$ & 12.68\% $\color{green} \uparrow$ & 0.36\% $\color{red} \downarrow$ & \vline & 21.25\% $\color{red} \downarrow$ & 20.20\% $\color{green} \uparrow$ & 0.48\% $\color{red} \downarrow$
       \\
    \hline
    B & [0.91; 0.95]\% & [0.51; 0.55]\% & [0.04; 0.05]\% & \vline & [0.04; 0.04]\% & [0.04; 0.04]\% & [0.00; 0.00]\% \\
       & 1.04 \% $\color{green} \uparrow$ & 0.42\% $\color{red} \downarrow$ & 0.03\% $\color{red} \downarrow$ & \vline & 0.00\% $\color{red} \downarrow$ & 0.04\%   & 0.00\%   \\
    \hline
 
    NAS & W & A & B & \vline & W & A & B \\
    \hline
    W & [79.63; 79.75]\% & [8.24; 8.34]\% & [1.40; 1.44]\% & \vline & [73.89; 74.29]\% & [3.48; 3.72]\% & [1.17; 1.28]\% \\
    	& 79.35\% $\color{red} \downarrow$ & 8.54\% $\color{green} \uparrow$ & 1.41\%  & \vline & 74.31\% $\color{green} \uparrow$ & 3.47\% $\color{red} \downarrow$ & 0.69 \% $\color{red} \downarrow$ \\
    \hline
    A & [2.94; 3.02]\% & [0.31; 0.38]\% & [0.11; 0.16]\% & \vline & [10.13; 10.46]\% & [0.81; 1.03]\% & [0.69; 0.69]\% \\
       & 3.24 \% $\color{green} \uparrow$ & 0.11\% $\color{red} \downarrow$ & 0.00\% $\color{red} \downarrow$ & \vline & 9.72\% $\color{red} \downarrow$ & 0.69\% $\color{red} \downarrow$ & 0.69\%  \\
    \hline
    B & [2.39; 2.46]\% & [0.26; 0.32]\% & [0.10; 0.12]\% & \vline & [0.00; 0.00]\% & [0.00; 0.00]\% & [0.00; 0.00]\% \\
       & 2.59 \% $\color{green} \uparrow$ & 0.22\% $\color{red} \downarrow$ & 0.00\% $\color{red} \downarrow$ & \vline & 0.00 \%  & 0.00\%   & 0.00\%  \\
    \hline
                 
  \end{tabular}
\end{table}

\end{document}